\documentclass[12pt]{article}
\pdfoutput=1
\usepackage{graphicx}
\usepackage{epstopdf}
\usepackage{amsmath}
\usepackage{amsfonts}
\usepackage{amssymb}
\usepackage{color}
\usepackage{mathrsfs}
    \usepackage{dsfont}
    
\newcommand{\be}{\begin{equation}}
\newcommand{\ee}{\end{equation}}
\newcommand{\bea}{\begin{eqnarray}}
\newcommand{\eea}{\end{eqnarray}}
\newcommand{\barr}{\begin{array}}
\newcommand{\earr}{\end{array}}

\def\beq{\begin{equation}}
\def\eeq{\end{equation}}
\def\be{\begin{equation}}
\def\ee{\end{equation}}
\def\bea{\begin{eqnarray}}
\def\eea{\end{eqnarray}}
\def\d{{\partial}}

\def\mpl{M_{\rm Pl}}

\def\zl{{\tilde\zeta_L}}
\def\dzl{{\dot{\tilde\zeta}_L}}
\def\dzln{\dot{\tilde \zeta}^{(n)}_L}
\def\sl{{S_L^{(\zl,n)}}}
\def\kl{k_{\rm low}}

\begin{document}

%\begin{titlepage}

\setcounter{page}{1} \baselineskip=15.5pt \thispagestyle{empty}

\begin{flushright}
%hep-th/yymmnnn\\
\end{flushright}
%\vfil

\begin{center}

{\Large \bf The constancy of $\zeta$\\[0.3cm] in single-clock Inflation at all loops}
\\[0.7cm]
{\large Leonardo Senatore${}^{1,2,3}$  and Matias Zaldarriaga${}^4$}
\\[0.7cm]
%\vspace{.7cm}
%\vspace{.3cm}
{\normalsize { $^{1}$ CERN, Theory Division, 1211 Geneva 23, Switzerland}}\\
\vspace{.3cm}

{\normalsize { $^{2}$ Stanford Institute for Theoretical Physics and Physics Department,\\
 Stanford University, Stanford, CA 94306}}\\
\vspace{.3cm}

{\normalsize {$^{3}$ Kavli Institute for Particle Astrophysics and Cosmology,\\
 SLAC and Stanford University, Menlo Park, CA 94025}}\\
\vspace{.3cm}

{\normalsize {  $^{4}$ School of Natural Sciences, Institute for Advanced Study, \\Olden Lane, 
Princeton, NJ 08540, USA}}\\
\vspace{.3cm}

\end{center}

\vspace{.8cm}

\hrule \vspace{0.3cm}
{\small  \noindent \textbf{Abstract} \\[0.3cm]
\noindent 
Studying loop corrections to inflationary perturbations, with particular emphasis on infrared factors, is important to understand the consistency of the inflationary theory, its predictivity and to establish the existence of the slow-roll eternal inflation phenomena and its recently found volume bound. In this paper we show  that $\zeta$-correlators are time-independent at large distances at all-loop level in single clock inflation. We write the $n$-th order correlators of $\dot\zeta$ as the time-integral of Green's functions times the correlators of local sources that are function of the lower order fluctuations. The Green's functions are such that  only non-vanishing correlators of the sources at late times can lead to non-vanishing correlators for~$\dot\zeta$ at long distances. When the sources are connected by high wavenumber modes, the correlator is peaked at short distances, and these diagrams cannot lead to a time-dependence by simple  diff. invariance arguments.  When the sources are connected by  long wavenumber modes one can use similar arguments once the constancy of $\zeta$ at lower orders was established. Therefore the conservation of $\zeta$ at a given order follows from the conservation of $\zeta$ at the lower orders. Since at tree-level $\zeta$ is constant, this implies constancy at all-loops by induction.
\vspace{0.3cm}
\hrule
%\vfil
%\begin{flushleft}
%\today
%March 20, 2008
%\end{flushleft}

%\end{titlepage}

%\newpage
%\tableofcontents
%\newpage

 \section{Introduction and Summary}
 
 This is the sequel to a series of recent papers~\cite{Senatore:2009cf,Senatore:2012nq,Pimentel:2012tw} that we wrote studying loop corrections to inflationary observables, with the purpose of understanding possible large infrared effects that might arise when considering the theory of inflation at quantum level. In the latest of these~\cite{Pimentel:2012tw}, we proved at one-loop level that in single clock inflation, that is inflationary theories where there is only one relevant degree of freedom, the correlation of the curvature perturbation $\zeta$ is time-independent on distances much longer than the horizon. Studying loop effect is in general important for the consistency of the theory at quantum level. However, the time-independence of $\zeta$ is very important also for the predictivity of inflation and for the existence of slow roll eternal inflation and its recently found universal volume bound~\cite{Creminelli:2008es,Dubovsky:2008rf,Dubovsky:2011uy} (see the introduction of~\cite{Pimentel:2012tw} where we recently discussed in detail about these points).
 
 The purpose of this paper is to extend the former calculation to all-loop orders.  We show the constancy of $\zeta$ iteratively in the number of perturbations. At linear level it is well known that $\zeta$ becomes time independent after horizon crossing. Then, assuming $\zeta$ is constant at a given perturbative order, we show that $\zeta$ is constant at the next perturbative order which provides a proof by induction. We achieve this in the following way.   The non-linear nature of the equations forces a coupling between different modes. The nature of the Green's function for the $\zeta$ fluctuations forces any interaction that happen in the past not to induce any time-dependence. Time dependence can only come from interactions that happened during the last Hubble time. Further, as it has been already shown in several ways (see for example~\cite{Lyth:2004gb,Langlois:2005qp,Maldacena:2002vr,Cheung:2007sv}), all interactions involving modes longer than the horizon are vanishingly small. This means that in the latest possible interaction at least two short wavelength fluctuations must be present, plus an additional arbitrary number of long wavelength fluctuations. This reduces the problem to computing correlation functions of local sources made of short-wavelength and long-wavelength fluctuations. The long wavelength fluctuations are lower order in the fluctuations, and so they can be assumed to be constant in time. However a constant-in-time long wavelength fluctuations is locally unobservable and therefore it cannot affect the short wavelength fluctuations. This means that the correlation of sources is dominated either by correlation of short modes at far away distances, that are too small to induce a time-dependence, or by correlations of derivatives of lower-order long-wavelength fluctuations, that decay in time. This implies that $\dot\zeta$ goes to zero at late times, proving the constancy of $\zeta$  at all loop order in single clock inflation. In other words, $\dot\zeta$ goes to zero in single clock inflation as an operator~\footnote{As described in~\cite{Pimentel:2012tw}, the non-linear dependence of $\zeta$ on lower order scalar or tensor fluctuations is parametrically different. This implies that the analysis can be done neglecting tensor modes, as we will do, though we expect that very similar arguments apply also when tensor modes are included.}. 
 
 \section{Local gauge for a long mode \label{sec:local_frame}}
 
It will turn out to be useful for the proof of the constancy of $\zeta$ to introduce the following set of coordinates, which was introduced for the first time in~\cite{Senatore:2012wy,Pimentel:2012tw} at linear level in the fluctuations.  Here we generalize them to the non-linear level. If we consider a set of $\zeta$ fluctuations $\zeta_L$ that involve, at the time of interest, long fluctuations characterized by  the highest wavenumber $k_{\rm low}$ already very outside of the horizon, we aim here to provide a change of coordinates valid in a local patch of volume $V$ much smaller than $k_{\rm low}^{-3}$ and much larger than $H^{-3}$, that takes us from the metric written in standard $\zeta$-gauge to a form that is locally of the form of a homogeneous anisotropic universe, up to corrections that go as $( k_{\rm low}^3 V)^{1/3}$. We wish to find a set of coordinates where this fact is manifest. Let us start in standard $\zeta$-gauge, defined by setting the scalar field to be uniform ($\delta\phi=0$ or the Goldstone boson $\pi=0$), and the metric to take the form
\be
ds^2=-N^2 dt^2+\sum_{ij} \hat h_{ij}\;\left(dx^i+N^i dt\right)\left(dx^j+N^j dt\right)\ ,
\ee
in the ADM parametrization. The spatial metric $\hat h_{ij}$ takes the following form
\be\label{eq:zeta-gauge}
\hat h_{ij}=\delta_{ij} \;e^{2\zeta}\ .
\ee
The variables $N$ and $N^i$ solve some constraint equations. For standard slow roll inflation (but the argument can be straightforwardly generalized to all models described by the EFT of Inflation~\cite{Cheung:2007st}), they are the following
\bea\label{eq:constrain}
&&\mpl^2\nabla_i\left[ N^{-1}\left(E^i{}_j+\delta^i{}_j E^l{}_l\right)\right]=0\ ,\\ \nonumber
&& \frac{\mpl^2}{2}\left(R^{(3)}-\frac{1}{N^2}\left(E^i{}_j E^j{}_i- E^i{}_i{}^2\right)\right)+\frac{\mpl^2}{N^2}\dot H-\mpl^2\left(3 H^2+\dot H\right)=0\ ,
\eea
where $R^{(3)}$ and $\nabla$ are taken with respect to the spatial metric $\hat h_{ij}$.
What makes the long wavelength limit in $\zeta$-gauge non-trivial is the fact that at linear order in Fourier space the solution for $N^i$ in the long wavelength limit reads 
\be
N^i\sim \frac{\kl^i}{\kl^2}\dot\zeta_L(k)\ ,
\ee
which does not have a nice $\kl\to 0$ limit for a generic time-dependence of $\zeta_L$. The same holds at higher order. If we go back to real space, this shows that at linear level $N^i\propto x^i$: $N^i$ becomes larger and larger at longer distances. The reason why this is not a problem is that $N^i$ is not really a physical observable: as we will show, what  matters is rather $\d_j N^i$.

In order to show this explicitly, we will solve the constraint equations (\ref{eq:constrain}) around the origin, at leading order in $(\kl^3 V)^{1/3}$. We can Taylor expand $\zeta_L(\vec x,t)=\zeta_L(0,t)+\left.\d_i\zeta\right|_0 x^i$. Because of rotational invariance, we can choose the gradient of $\zeta_L$ to point in the $z$ direction. It can be checked that, non-linearly in $\zeta_L$, but at leading order in $(\kl^3 V)^{1/3}$, we have
\bea
&& N=1+\frac{\dot\zeta_L}{H}+{\cal O}\left((\kl V)^{1/3}\right)\ , \\ \nonumber
&& N_i=N_{i,0}(t)-\delta_{i3}\frac{\dot H}{2 H^2}\,\frac{2 H+\dot \zeta_L}{H+\dot\zeta_L}\,\dot\zeta_L\; z + {\cal O}\left((\kl ^3 V)^{1/3}\right)\ ,
\eea
where all the $\zeta_L$ are evaluated at the origin and $N_{i,0}$ are three space-independent functions. For later convenience, let us define
\be
F(\dot\zeta_L)=\frac{\dot H}{2 H^2}\,\frac{2 H+\dot \zeta_L}{H+\dot\zeta_L}\,\dot\zeta_L\ .
\ee
As mentioned, the problematic term is $N_i$: it does not appear to go to zero as $\kl\to 0$, even after it has been acted upon by a spatial derivative. Notice however that this term diverges linearly in $z$, suggesting that the physically sensible quantity is rather $\d_j N_i$. Indeed, the structure of the constraint equations show that one can restrict oneself to solutions that diverge at most linearly in the spatial coordinated. This means that in order to find a reference frame where the universe looks locally manifestly like an homogeneous anisotropic universe, it is enough to find a coordinate transformation that set to zero locally $N_i$ and $\d_j N_i$, with the remaining fluctuations going to zero as $k_{\rm low}\to 0$, after have been acted upon by a spatial derivative.
We will work without assuming any specific time dependence for $\zeta_L$, so that the following change of coordinates holds also when one cannot use the solution to the linear equations for $\zeta_L$, as it happens for example when computing loop corrections to inflationary correlators~\cite{Pimentel:2012tw}. 

We are going to perform the following change of coordinates:
\be
\tilde x^i=M^i{}_j(t)\; x^j+C^i(t)\ , \qquad {\rm with}\qquad {\rm det}[M]\neq 0 \ ,
\ee
which keeps the inflaton fluctuations zero. 
Under this change of coordinates, and expanding around the origin, the metric takes the form
 \bea \nonumber
&& ds^2=-N^2 dt^2+e^{2\zeta(\vec x(\tilde x,t),t)}\delta_{ij}\left[M^i{}_l\, d\tilde x^l+\left(\dot M^i{}_l \tilde x^l+\dot C^i+\left.N^i\right|_0+\left.\d_l N^i\right|_0\left(M^l{}_m\, \tilde x^m+C^l\right)\right)dt\right]\\
 &&\quad\qquad\qquad\qquad\qquad\times \left[M^j{}_n\, d\tilde x^n+\left(\dot M^j{}_n \tilde x^n+\dot C^j+\left.N^j\right|_0+\left.\d_n N^j\right|_0\left(M^n{}_p\, \tilde x^p+C^n\right)\right)dt\right]\ .
 \eea
To bring the metric in our desired form, we therefore need to solve the following two equations:
 \bea
&& \dot C^i+ \left.\d_l N^i\right|_0 C^l+\left.N^i\right|_0=0\ ,
 \\ \nonumber
&&\left(\dot M^i{}_l+\left.\d_n N^i\right|_0M^n{}_l\right) \tilde x^l=0 \ .
 \eea
   The generic solution reads
   \bea
   && C^i(t)=-\left[U(t)^{-1}\right]^i{}_j\left[\int^t_{t_2} dt' U(t')^j{}_l \left.N^l\right|_0(t')\right] + \left[U(t)^{-1}\right]^i{}_j C_1^i\ ,\\ \nonumber
   && M(t)^i{}_j=\left[U(t)^{-1}\right]^i{}_l\; M_1^l{}_j\ ,
   \eea   
where
\be
U=T e^{\int^t_{t_1} dt'\;\left.\d N\right|_0(t')} \ ,
\ee
with $T$ staying for time-ordering,  $t_{1,2}, C_1, M_1^{i}{}_{j}$ representing integration constants, and $\left.\d N\right|_0$ is the matrix such that $[\left.\d N\right|_0]^i{}_j=\left.\d_j N^i\right|_0$.

After substitution, the metric takes the form
\be\label{eq:non_linear_change}
   ds^2=-N^2 dt^2+e^{2\zeta(\vec x(\tilde x,t),t)}\delta_{ij} M^i{}_l M^j{}_m\, d\tilde x^l d\tilde x^m\ .
 \ee
 Since $M$ contains only $\d N$ terms, the metric goes to zero as $\kl\to 0$, if acted upon by spatial derivatives, as we wished to show. 
 The local spatial metric becomes
 \be\label{eq:local_metric}
 g_{L,ij}^{\rm space}=a^2 e^{2 \tilde\zeta_L(t)\; \mathds{1}+\sum_{n=1,\ldots,5} {\tilde\lambda_{L,n}(t)}\; S_{n,ij}}
 \ee
 where $\tilde \zeta_L$ is the local $\zeta$ fluctuation, that multiplies the identity matrix $\mathds{1}$, and $\tilde\lambda_{L,n}$'s are the local anisotropy coefficients, represented as coefficients of five independent symmetric traceless matrixes $S_n$~\footnote{These are given by
 \bea
 &&S_{1,ij}=\delta_{(i2}\delta_{j)3}\ , \qquad  S_{2,ij}=\delta_{(i1}\delta_{j)3}\ , \qquad  S_{3,ij}=\delta_{(i1}\delta_{j)2}\ , \\ \nonumber
 &&S_{4,ij}=+\frac{1}{3}\delta_{i1}\delta_{j1}-\frac{2}{3} \delta_{i2}\delta_{j2}+\frac{1}{3} \delta_{i3}\delta_{j3}\ , \qquad S_{5,ij}=+\frac{1}{3}\delta_{i1}\delta_{j1}+\frac{1}{3} \delta_{i2}\delta_{j2}-\frac{2}{3} \delta_{i3}\delta_{j3}\ ,
 \eea
 where $(i,j)$ means symmetrization with respect to the $i,j$ indexes. }.

Our choice to align the gradient of $\zeta_L$ with the $z$ direction simplifies formulae remarkably. We have:
\be
\left[\left.\d N\right|_0\right]_{ij}=\delta_{i3}\delta_{j3} F(\dot\zeta_L)\ ,\qquad 
U=
\left(
\begin{array}{ccc}
1 &0  &0 \\
0 &1 &0 \\
0 &0  & e^{\int^t_{t_1} dt' \;F(\dot\zeta_L(t'))}\ 
\end{array} \right)\ ,
\ee
In the local frame the fluctuations, represented by a tilde, are given in terms of $\zeta$ by the following relation
\bea\label{eq:detailed_local}
&& \zl(\vec {\tilde x},t)=\zeta_L(\vec x(\vec{\tilde x}),t)+\frac{1}{3}\int^t_{t_1}dt'\;\frac{\dot H}{2 H^2}\frac{2 H+\dot \zeta_L(\vec 0,t')}{H+\dot\zeta_L(\vec 0,t')}\,\dot\zeta_L(\vec 0,t')\ ,\\ \nonumber
&& \tilde\lambda_{L,5}=-\int^t_{t_1}dt'\; \frac{\dot H}{2 H^2}\frac{2 H+\dot \zeta_L(\vec 0,t')}{H+\dot\zeta_L(\vec 0,t')}\,\dot\zeta_L(\vec 0,t')\ .
\eea
All the remaining $\lambda_{L,n}$ vanish. For this reason, in the rest of the paper, we will call $\tilde\lambda_{L,5}$ as simply~$\tilde\lambda_{L} $~\footnote{Later in the proof of the constancy of $\zeta$, we will need to consider configurations where there is a generic lower-order long-wavelength configuration sourcing a long wavelength fluctuations. A priori, these might have different gradient directions. However, since we work at linear order in the gradients, rotational invariance implies they must have the same direction. Alternatively, one could imagine to perform a rotation to bring the grandient in the $z$-direction, obtain the local frame in this simple coordinates, and then perform the inverse rotation. In this way, one would obtain all the $\lambda_{L,n}$.  In any event it is straightforward to use the generic $\lambda_{L,n}$ from the get-go.}. Notice that under this change of coordinates only the long wavelength $\zeta_L$ is transformed. Short wavelength fluctuations $\tilde\zeta_S$ transform as a scalar. The long wavelength metric takes locally the following form
\be\label{eq:local_metric}
g_{L,\mu\nu}=-N^2 dt^2+g^{{\rm space}}_{L,ij}\;dx^i dx^j\ ,
\ee
with the spatial metric defined in (\ref{eq:local_metric}) and given here in matrix form
\be
g_{L}^{\rm space}(t)=a(t)^2{\left(
\begin{array}{ccc}
e^{2\left(\zl(t)+\frac{1}{3}\tilde\lambda_{L}(t)\right)}& 0 & 0 \\
0& e^{2\left(\zl(t) +\frac{1}{3}\tilde\lambda_{L}(t)\right)} & \\
0 & 0 & e^{2\left(\zl(t) -\frac{2}{3}\tilde\lambda_{L}(t)\right)}
\end{array} \right)}\ ,
\ee
and 
\be
N(t)=1+\frac{\dot\zeta_L\left(\dot{\tilde\zeta}_L(t)\right)}{H}
\ee
Notice that ${\rm det}[g_L]=N^2 a^6 e^{6\zl}$. The long wavelength mode has been locally reabsorbed in an homogenous anisotropic background up to corrections that are manifestly subleading in $(\kl^3 V)^{1/3}\to~0$ limit. Short wavelength fluctuations live in this very smooth universe.

 \section{Proof of the conservation of $\zeta$ to all loops}
 
 Let us consider a long wavelength fluctuations $\zeta_L(\vec x,t)$, defined as
 \be\label{eq:global_zeta_long}
 \zeta_L(\vec x,t)=\int d^3k\; \Theta(k_{\rm low}-k)\; e^{i\vec k\cdot \vec x}\zeta_{\vec k}(t)\ ,
 \ee
 where $k_{\rm low}$ is some comoving momentum.  We are interested in proving the constancy of correlation functions of $\zeta_L(\vec x,t)$ once all the comoving modes that are part of its spectrum are outside of the horizon. As we solve iteratively in the interactions the equation of motions for $\zeta_L$, we need to insert interactions at some specific times and spatial locations, and sum over these. In~\cite{Senatore:2009cf,Pimentel:2012tw}, we have shown how solving iteratively in the interactions the equations of motion and then taking correlation functions of the non-linear solutions is equivalent to computing loop corrections. Given a certain solution with the insertion of a given number of interactions, there will be one interaction that has been inserted last. Let us consider it. It is sufficient to work with the tilde variables of the local frame defined in Sec.~\ref{sec:local_frame}. We can write the equation of motion for $\zl$ and the anisotropy coefficient  ${\tilde\lambda}_{L}$ in the following way
 \bea\label{eq:first}
&& \left(\d_t+3 H \right) \dot{\zeta}_L^{(n)}(\vec x,t)= \left(\d_t+3 H \right)\left(\frac{3 H^2}{3 H^2+\dot H}\;\dot{\tilde \zeta}^{(n)}_L\right)=\sl(\vec x,t)\ , \\ \nonumber
&&  \left(\d_t+3 H \right) \dot{\tilde\lambda}_{L}^{(n)}(\vec x, t)=S_L^{(\tilde\lambda_{L},n)}(\vec x,t)\ ,  
 \eea
where $\sl$ and $S_L^{(\tilde\lambda_{L},n)}$ are some source that depend in a local way on $\tilde\zeta_L$, $\tilde\lambda_{L}$, $\delta {\tilde N}_L$, and their derivatives, where $\tilde\zeta_L,\delta\tilde{ N}_L,\tilde \lambda_{L}$ are respectively $\zeta$, the time-shift and  the anisotropy coefficient of the local space-time described in Sec.~\ref{sec:local_frame}. $S_L$ additionally depends in a more unconstrained way on the additional short wavelength fluctuations. Let us explain how we  obtain the above equation: 
\begin{enumerate}
\item First of all, notice that the differential operator acting on $\zeta_L$ on the left is very similar to the differential operator that acts on linear fluctuations in the standard $\zeta$-gauge, apart for a $\nabla^2/a^2$ term which is missing. The gradient is indeed negligible because we are interested in proving the constancy of $\zeta_L$ when $k_{\rm low}$ is already very outside of the horizon. Since, as we will see next, both the $\zeta$-Green's function and $\tilde \zeta$-Green's  function of the long mode makes any solution that is sourced by a source in the past approach a constant, it is only the interactions that are happening at the latest possible time that could lead to a time-dependence. This means that in~(\ref{eq:first}) we can restrict ourselves to times when $k_{\rm low}$ is outside of the horizon, and so the term $\nabla^2\zeta_L$ is negligible.

\item The right hand side of (\ref{eq:first}) contains the source term $S_L$. As we said, this term is a local function of $\tilde\zeta_L$, $\tilde \lambda_{L}$, $\delta \tilde N_L$ and their derivatives, and of the short wavelength fluctuations in a more unconstrained way. In general, each of the fluctuations in $S_L$ will be either a free field, or the results of some interactions. Each interaction is associated to a vertex with some fluctuations, and so it is useful to express the perturbative series in terms of number of fluctuations involved, rather than in the number of vertices included. This is equivalent to the counting of loops~\footnote{The same phenomenon happens in standard quantum field theory. For example, in the correction to the propagator at first order in $\hbar$,  diagrams with either one quartic or two cubic vertices contribute. In standard quantum field theory, indeed, fluctuations are of order $\hbar$.}.  Let us call $\tilde\zeta^{(n)}$ as the solution to the perturbative equations of order~$n$ in the fluctuations. If we wish to source $\tilde\zeta_L^{(n)}$, the sum of the orders of all the $\tilde\zeta_L$, $\tilde \lambda_{L}$, $\delta \tilde N_L$ and the short fluctuations in $S_L$ must be equal to $n$. In particular, notice that the fluctuations $\tilde\zeta_L,\, \tilde \lambda_{L}$ and $\delta \tilde N_L$ appearing in $S_L$ are of lower order in the fluctuations. The  structure of the equation suites itself therefore to an iterative solution. We will prove that the assumption that the lower order $\tilde\zeta_L$ is constant implies that the next higher order $\tilde\zeta_L$ is constant as well. Since at tree level $\tilde\zeta_L$ is constant, the argument provides a proof to all orders~\footnote{When $\zeta$ is not constant at tree-level or alternatively when one cannot go to standard $\zeta$-gauge as defined in~(\ref{eq:zeta-gauge}), as for example in~\cite{Endlich:2012pz}, our proof does not apply.}. Notice immediately that assuming the lower order $\tilde\zeta_L$ to be constant simplifies the structure of the equation, as if $\tilde\zeta_L$ is constant, $\delta \tilde N_L$ must vanish and $\tilde \lambda_{L}$ must go to a constant value.

\item How can we guarantee that $S_L$ is spatially local? Once a mode is much longer than the Horizon, the Green's function associated to the differential operator in (\ref{eq:first}) is local in space, as we will show next in (\ref{eq:green}).  This means that a long wavelength mode at one location is sourced  only by a source at the same location. This shows that the interactions of the long mode with the other modes can be effectively described in the local anisotropic frame introduced in Sec.~\ref{sec:local_frame}, and we can use directly the  variables with tildes. Notice that the constancy of $\zl,\tilde\lambda_{L,n}$ implies the constancy of $\zeta_L$ and viceversa. In this frame it is manifest that the dependence of $S_L$ on the long-wavelength fields is manifestly local. In more detail it follows from the fact that in the constraint equations for the short-wavelength $\delta\tilde N_S$ and $\tilde N^i_S$ there cannot be spatial derivatives acting on $\tilde\zeta_L$ and $\tilde \lambda_{L}$ as they can be taken as homogeneous. Corrections to this are explicitly suppressed by powers of~$k_{\rm low}/(a H)$. Notice that this is not the case for standard $\zeta$-gauge, where  operators such as $\d_i/\d^2$ acting on long-wavelength fields, appear in the source.

 \end{enumerate}
 
Since we are interested in $\dzl$, the solution to (\ref{eq:first}) can be written directly:
\be
\dzln(x,t) =  \int_V d^4x^\prime\; {\cal D}_{x'}\,\tilde G_L(x,x^\prime); \sl(x^\prime)\ ,
\ee
 where $\tilde G_L(x,x')$ is the $\dzl$-Green's function for the long mode
  \be\label{eq:green}
\tilde G_L(x,x')=\delta^{(3)}(x-x')\;G_L(t,t')\ ,\qquad G_L(t,t')=\frac{3 H^2+\dot H}{H^2}\frac{e^{-3H (t-t')}}{3 H}\ ,
 \ee
 in the slow roll approximation, and ${\cal D}_{x'}$ is the $x'$-differential operator that acts on $\zeta$ in the vertex that gives rise to $S_L$. Since we can neglect gradients of the long mode, the Green's function is local in real space. We have
\be\label{eq:solution}
\dzln(t) = \int dt^\prime\; {\cal D}_{t'}\, G_L(t,t^\prime)\; \sl(t^\prime) \ .
\ee
As promised, for $t\gg t'$, $G_L$ tends to a constant, which means that in order for a time-dependence to be induced, we need to concentrate on $t'\sim t$. All of the above applies identically to the sourcing of $\tilde\lambda_{L}$ after replacing $\zl$ with $\tilde\lambda_{L}$ apart from an irrelevant multiplicative constant in the definition of the Green's function, that is time-idependent at leading order in slow-roll. Slow-roll corrections are irrelevant for the purpose of proving the constancy of $\zeta_L$.  

Since generic $n$-point functions are formed by decomposing them as product of connected correlation functions involving a smaller number of fluctuations, we can concentrate on connected correlation functions, that are therefore simply related to connected correlation functions of $S_L$. Since the case of correlation functions involving also $\tilde\lambda_{L}$ follows virtually unaltered from the case of the correlation function of $\zl$, we quote explicit results only for $\zl$. We concentrate first on the two point function. We summarize the following points:
 \begin{enumerate}
\item Some of the operators contributing to at least one of the $\tilde \zeta_L$ must involve short modes, as otherwise all interactions vanish. 
\item We are interested in contributions at late times because the form of the Green's function makes~$\tilde\zeta_L$ go to a constant if the final interaction is in the distant past. 
\item Any vertex connected to the $x^\prime$ by a short mode must be within a space-time distance of order Hubble at most. This is because short modes need to be oscillating and they become uncorrelated on larger distances or longer times~\footnote{This is quite intuitive, but let us give a detailed explanation. The fact that they become uncorrelated in time can be quickly understood by realizing that one can rotate the contour of integration in such a way that all modes inside the horizon are exponentially damped as $e^{k \eta}$. Given that the last vertex has to be at about horizon crossing, the earlier one cannot be further in the past than about one Hubble time. The fact that short modes $k_S$ with physical wavenumber greater or equal than Hubble do not contribute to the correlation at long physical distances $r_L(t)\gg H^{-1}$ can be seen by a simple Fourier space argument: the contribution of an $e$-folding of modes to the correlation at a physical distance $r_L(t)$ scales as $\sin(r_L\; k_S/a)/(r_L\; k_S/a)$. Since $k_S/a\gtrsim H$, this contribution vanish as we take $r_L(t)\gg H^{-1}$.}.

\end{enumerate}

Let us consider the correlator of $\sl$, and first consider taking expectation value only over the short modes. This expectation value should be thought of as taken on a diff-invariant regularized theory, for example in dimensional regularization, with the counter terms provided naturally by the operators present in the effective theory of single clock inflation~\cite{Cheung:2007st}~\footnote{An explicit regularization and renormalization using the operator structure of effective field theory of single clock inflation was made in~\cite{Senatore:2009cf} and~\cite{Pimentel:2012tw}, where the symmetry constraints provided by the effective theory turned out to be extremely useful to simplify the calculations.}. The general structure takes the form:
\bea\label{eq:short_average} \nonumber
&&\langle\sl(x_1) \sl(x_2) \rangle_{\rm short}  \\ \nonumber
&&\qquad\qquad=\frac{\delta^{(4)}\left(x_1-x_2\right)}{a^3 e^{3\zl}} \; F_A\left(t,\dzl(x_1),\ldots\right) 
%\qquad\qquad\qquad\qquad\qquad\qquad\qquad   %\\ \nonumber
+F_B\left(t,\dzl(x_1),\ldots\right)F_B\left(t,\dzl(x_2),\ldots\right)\ . \\ 
\eea
where $\ldots$ represents $\dot{\tilde\lambda}_{L}$, $\delta \tilde N_L$ and higher time derivatives of all these functions evaluated at the same location. The first term accounts for cases where the vertices around $x_1$ and $x_2$ are connected by high momentum lines (which there must be at least two by momentum conservation). Note that if there are short momentum lines connecting the vertices they have to be close both in space and in time. We call these Short-Connected ($SC$) diagrams and they  reduce in the two point function at 1-loop to the cut-in-the-middle diagrams discussed in~\cite{Senatore:2009cf,Pimentel:2012tw}. Since at large distances these modes become uncorrelated, the contribution is peaked at the same spatial location by a simple $1/N$ argument (see Fig.~\ref{fig:loop_CIM}). We approximated the spatial dependence as a $\delta$-function, though it should be really thought of as a function that has support within a distance of order Hubble. For the purposes of proving the constancy of $\zeta$ at long wavelength, this difference is irrelevant. 

We made the $\delta$-function a scalar function of the local coordinates. This is why there is a factor of the determinant of the local metric $g_{L,\mu\nu}$, which is given by (\ref{eq:local_metric}). The reason why we made the $\delta$-function a scalar can be  understood by noticing that there is a residual gauge-transformation that keeps us in the local homogeneous anisotropic coordinates. This is given by a time-independent rescaling of the coordinates
\be\label{eq:residual_gauge}
x^i\rightarrow M^i{}_j x^j \ .
\ee
This change of coordinates does not affect $\dzl$ and $\dot{\tilde\lambda}_{L}$, but does change $\zl$ and $\tilde\lambda_{L}$, and it can be checked that all these variables can be changed  independently. In particular, for the choice of coordinates in which the gradient of $\zeta_L$  is on the $z$ axis, so that only $\tilde\lambda_{L,5}$ is non-vanishing, we have the two independent transformations:
\bea
&&x^i\rightarrow e^\alpha x^i \ ,\qquad \tilde\zeta_L\rightarrow\tilde\zeta_L-\alpha\ , \\ \nonumber
&&x^i=\left[e^{S_5\beta}\right]^i{}_j\;x^j=\left(
\begin{array}{ccc}
e^{\beta/3} &0  &0 \\
0 &e^{\beta/3} &0 \\
0 &0  & e^{-2\beta/3}\ 
\end{array} \right)_{ij} x^j\ , \qquad  \tilde\lambda_{L,5}\to\tilde\lambda_{L,5}+\beta\ .
\eea
The right hand side of (\ref{eq:short_average}) must therefore be invariant under the residual gauge-transformation. This implies that every $x$-dependence must be accompanied with a contraction with the local metric. The factor of $a^3$ in the local metric must be there because we can choose to re-absorb a isotropic rescaling of the coordinates with a rescaling of~$a$. Additionally, this implies that $F_A$ cannot depend on $\zl$ and $\tilde \lambda_{L}$, but only on $\delta \tilde N_L,\,\dzl,\ldots\ldots$, all of these vanishing as some power of $k_{\rm low}/(a H)$. We stress that at diagrammatic level the absence of $\zl$ in the expectation value of the short modes appears in general as the result of a cancellation among many different diagrams. The complex structure and detailed physical reasons of how this cancellation occurs was studied in~\cite{Pimentel:2012tw}.

The second term represents the contribution from diagrams where $x_1$ and $x_2$ are connected by low momentum lines (see Fig.~\ref{fig:loop_CIS}). We call these Long-Connected ($LC$) diagrams. At 1-loop in the 2-point function calculation, they reduce to the a cut-in-the-side diagrams discussed in~\cite{Senatore:2009cf,Pimentel:2012tw}. $F_B$ must therefore vanish if any of the long wavelength fluctuations vanish. The same gauge transformation as in (\ref{eq:residual_gauge}) implies that $F_B$ cannot depend on $\zl$ and $\tilde\lambda_{L}$ but only on its derivatives, which vanish as powers of $k_{\rm low}/(a H)$. 

We have taken the dependence of the functions $F_A$ and $F_B$ on  $\dzl,\;\delta \tilde N_L,\ldots$ as local in time. This is only an approximation. The iterative argument implies that they decay in time with a time scale of order Hubble. However, it should be noticed that, because of the $i\epsilon$ prescription, the time integrals have support of order one Hubble time, and, for the purposes of proving the constancy of $\zl$, we can approximate the dependence as local in time. 
$F_A$ and $F_B$ can also depend on time, but in a way that is proportional to the slow-roll parameters for example through the slow changes in $H$. Also note that these terms include connections through long modes that `are generated'  in other vertices that might be outside the volumes around $x_1$ and $x_2$ and are connected to both via long modes. The $\zl$ appearing inside the $F$s account for the entire $\tilde \zeta_L$ which is constant over each $V$ region (see Fig.~\ref{fig:consistency_external}).

\begin{figure}[h]
\begin{center}
\includegraphics[width=9cm]{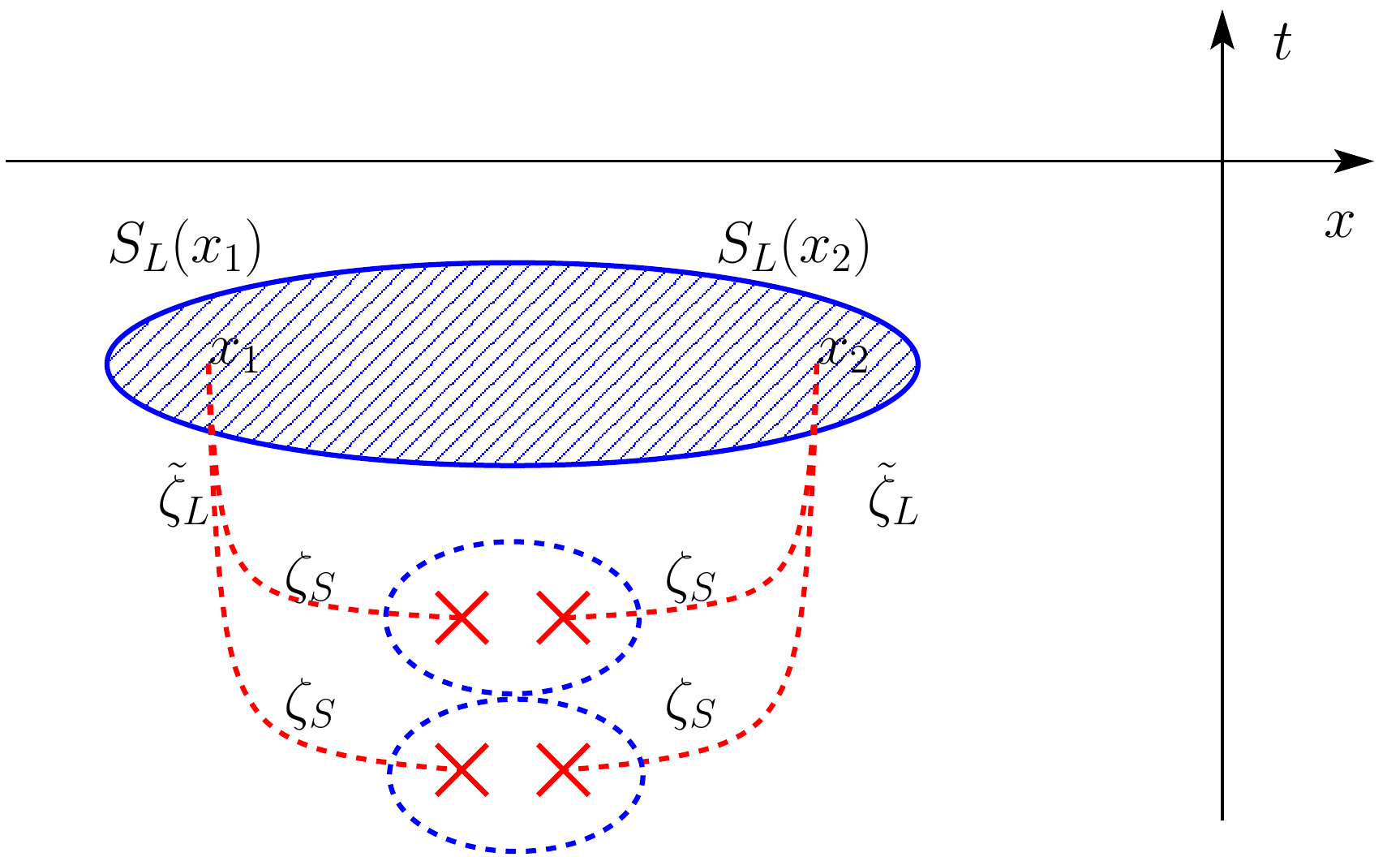}
\caption{\label{fig:loop_CIM} \small\it Example of a $SC$ (short-connected) diagram. Red-dashed lines ended by a red-cross represent free modes, and two crosses encircled by a blue-dashed ellipse represent a correlation function between those two free modes. The dashed-filled ellipse represents possible additional correlation functions. This is a $SC$ diagram as, among the various correlation functions present in the diagram, there are two involving two short modes at different locations $x_1$ and $x_2$ that are contracted among each other. If the distance between $x_1$ and $x_2$ is much longer than the wavelength and frequency of the two modes, the contribution becomes vanishingly small. The points $x_1$ and $x_2$ have therefore to be very close to each other. This diagram contributes to the first term of (\ref{eq:short_average}).}
\end{center}
\end{figure}

\begin{figure}[h]
\begin{center}
\includegraphics[width=9cm]{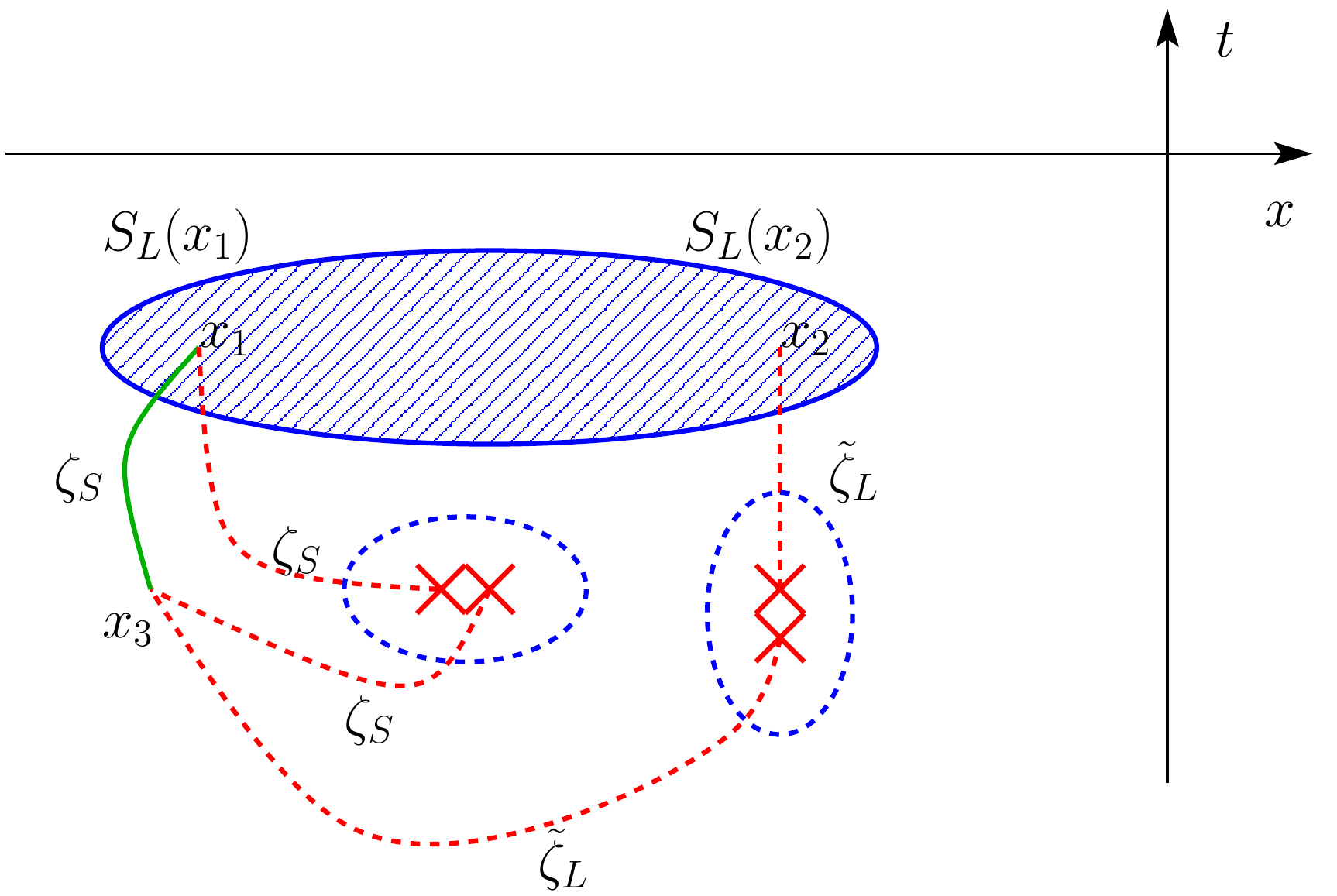}
\caption{\label{fig:loop_CIS} \small\it Example of a $LC$ (long-connected) diagram. Green line represent Green's function. The dashed-filled ellipse contains possible additional correlations that connect $x_1$ and $x_2$ only though long modes.  This is a $LC$ diagram as $x_1$ and $x_2$ are connected only by low momentum lines. Notice that $x_3$ must be very close to $x_1$ because it is connected to $x_1$ by a correlation of short modes. This diagram contributes to the second term of (\ref{eq:short_average}).}
\end{center}
\end{figure}

\begin{figure}[h]
\begin{center}
\includegraphics[width=9cm]{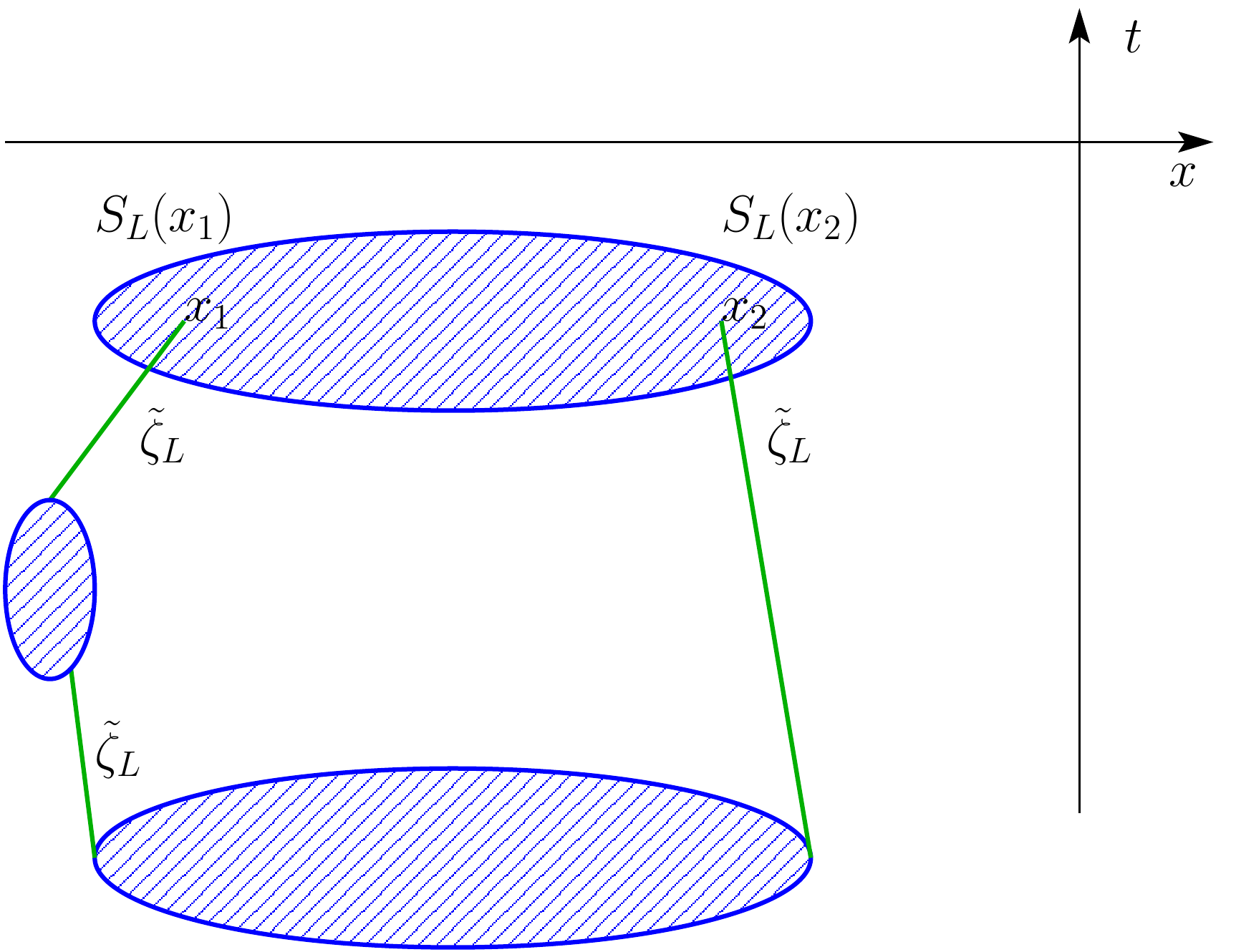}
\caption{\label{fig:consistency_external} \small\it Example of a diagram where the $\zeta_L$ dependence of the functions $F_A$ and $F_B$ appearing in the average over short modes of $\langle\sl(x_1) \sl(x_2) \rangle_{\rm short}$ (top shaded region) is generated by some interaction in the past (shaded regions).  If the ellipses enclosing $x_1$ and $x_2$ contains a correlation function made of short modes  that connect $x_1$ and $x_2$ we obtain a $SC$ diagram, and it contributes to the first term of~(\ref{eq:short_average}). If instead $x_1$ and $x_2$ are connected only by long modes,  we have a $LC$ diagram, and a contribution to the second term of~(\ref{eq:short_average}). After taking expectation value over the long modes, we obtain a contribution respectively to the first and the second line of (\ref{eq:long_short_average}).}
\end{center}
\end{figure}

Let us know take the average over the long modes:
\bea\label{eq:long_short_average} \nonumber
&&\langle\sl(x_1) \sl(x_2) \rangle_{\rm short\,\&\, long} = {\delta^{(4)}(x_1-x_2)\over a(t_1)^3} \langle e^{-3\zl}F_A(t,\dzl,\ldots)\rangle
\\ &&\quad \qquad \qquad\qquad \qquad\qquad\qquad 
+ \left\{F_{B1}(t_1)F_{B1}(t_2) \langle\dzl(x_1)\dzl(x_2)\rangle+ \cdots \right\} ,
\eea
where $F_{B1}=\left.\d F_B/\d\dzl\right|_0$. Let us consider the expectation value in the first line. Since the lower order $\dzl$ vanish, this is dominated the the expectation value of the determinant of the metric. 
We have 
\be\label{eq:expt}
\langle e^{-3\zl}\rangle\simeq 1+\frac{9}{2}\langle\zl^2\rangle+\ldots \ ,
\ee
where we have expanded the argument of the expectation value to the leading order. Since $\zl$ is made of lower order modes with wavenumber up to $k_{\rm low}$, all of which are constant, this expectation value is non-perturbatively time- (and space-) independent. The neglected terms decay in time as powers of $k_{\rm low}/(a H)$. 
Analogously the expectation value on the second line decays in time as it is made up of lower order $\dzl$  that are assumed to be decaying as some power of $k_{\rm low}/(a H)$ by the iterative argument. The same is true for the higher derivative terms. The details of how the decay happens might be dependent on which specific vertex we are considering. In particular, we expect generically the suppression to go as $(k_{\rm low}/(a H))^2$, but note that one factor of $({k_{\rm low} /( a H)})$ suffices to make the $\zl$ correlation time independent. 
The $\dzl$ dependence in the first line originates for example  from short-connected diagrams that are connected additionally by long modes (see Fig.~\ref{fig:consistency_external}). 
Additionally, notice that if there are short vertices around both $x_1$ and $x_2$ which are being affected by the $\zl$ in the correlation (as opposed of the the line being connected directly to the external $\zl$) one should be paying a price in {\it each} vertex.

Let us now compute the two point function of $\dzl$. We have
\bea\label{eq:final}
&&\langle\dzl(x_1)\dzl(x_2)\rangle=\int^t dt_1\int^t dt_2\ e^{-3 H (t-t_1)}  e^{-3 H (t-t_2)} \\ \nonumber
&&\qquad\qquad\qquad\qquad\qquad\left\{\frac{ \delta^{(4)}(x_1-x_2)}{a(t_1)^3} \left[\tilde  F_{A0}(t_1) + \left({k_{\rm low} \over a(t_1) H}\right) \tilde F_{A1}(t_1) + \cdots \right] \right. \nonumber \\ \nonumber
&& \qquad\qquad\qquad\qquad\qquad +\left. \left[\tilde  F_{B1}(t_1) \tilde  F_{B1}(t_2) \left.\left(\frac{\d_{x_1-x_2} }{ a H}\right)^2\right|_{{\rm Min}[t_1,t_2]} C(x_1-x_2)+ \cdots \right]\right\}\ .
\eea
where $C(x_1-x_2)$ is the lower-order correlation function of $\zl$, and we have rearranged the terms according to their $\d_{x_1-x_2}/(aH)$ suppression, naming the coefficients of each term as $\tilde F_{A\dots}, \tilde  F_{B\dots},\ldots\ $. We have used the slow roll approximation for the Green's function of $\dot\zeta$: $G_L(t,t')\sim {\rm Exp}(-3H(t-t'))$, the subleading corrections being irrelevant.
Notice that since $C(x_1-x_2)$ is approximately scale invariant, its spatial derivative is peaked at short distances.

 The time integrals clearly go to zero at late times~\footnote{
 It might be useful to notice that
 \bea \nonumber
&& \int^t_{t_{in}} dt_1\;e^{-6 H(t-t_1)} e^{-3 H t_1}=\frac{e^{-3 H t}}{3\,H} \left(1-e^{-3 H (t-t_{in})}\right)\  , \\ \nonumber
&&\int^t_{t_{in}} dt_1\; \int^{t_{1}}_{t_{in}} dt_2\; e^{-3 H(t-t_1)}e^{3 H(t-t_2)} e^{-2 H t_2}= \frac{e^{-2 Ht}}{12\, H^2}\left(3- 4\, e^{-H(t-t_{in})}+e^{-4H(t-t_{in}) }\right)\ .
 \eea
 }. Notice that the spatial dependence of the correlation function consists of the sum of a $\delta$-function and a term that is peaked at short distances approximately as $1/x^2$.
  
The same behavior is true when we consider the contribution of $n$-point correlations involving $\dzl$ and/or $\dot{\tilde\lambda}_{L,n}$. This shows that the correlation functions of $\dzl$ and $\dot{\tilde\lambda}_{L,n}$ at a given order vanish once we take the lower order correlation functions to be time-independent. This proves by iteration that $\dzl$ and $\dot{\tilde\lambda}_{L,n}$ have vanishing correlation functions at all orders at late times. They vanish as operators.

Finally, let us consider the long wavelength $\dot\zeta_L$ defined in (\ref{eq:global_zeta_long}). As given by ($\ref{eq:non_linear_change}$) and (\ref{eq:detailed_local}) in Sec.~\ref{sec:local_frame}, the relationship between $\dot{\tilde\zeta}_L$ and $\dot{\tilde\lambda}_{L}$ is such that once $\dot{\tilde\zeta}_L$ and $\dot{\tilde\lambda}_{L}$ vanish so does $\dot\zeta_L$, implying that the correlation functions of $\dot\zeta_L$ vanish to all orders for $k_{\rm low}/(a H)\ll 1$. In particular, since for $k\leq k_{\rm low}$, $\zeta_L(k)=\zeta(k)$, with $\zeta(k)$ being the Fourier transform of the globally defined $\zeta$ fluctuation, the correlation functions of $\dot\zeta_k$ vanish at all orders for $k/(a H)\ll 1$, as we ultimately wished to show.

   We finally make a comment on the IR divergencies present in the above formula (\ref{eq:final}). Indeed notice that if we try to evaluate (\ref{eq:expt}) perturbatively, we obtain, at leading order,
  \be\label{eq:expt2}
  \langle e^{-3\zl}\rangle\simeq 1+\frac{9}{2}\langle\zl^2\rangle+\ldots \simeq 1+\frac{9}{2} \Delta_\zeta \log(k_{\rm low} L )+\ldots\ ,
  \ee
 where $L$ is the infrared comoving size of the box and where $\Delta_\zeta$ is the tree-level dimensionless power spectrum of the $\zeta$ fluctuations  
\be
\langle\zeta(\vec k,t\to\infty)\zeta(\vec k',t\to\infty)\rangle_{\rm tree-level}=(2\pi)^3\delta^{(3)}(\vec k+\vec k') \;\frac{\Delta_\zeta}{k^3}\ ,
\ee
where we have neglected the tilt. We see that there is an IR divergence as $L\to \infty$, hidden in (\ref{eq:final}) in the coefficients $\tilde F_{A\dots},\tilde F_{B\ldots},\ldots\ $. Such IR divergencies were found first in~\cite{Giddings:2010nc,Byrnes:2010yc} in the context of the long wavelength $\zeta$ 1-loop power spectrum, but they are present in every $\zeta$-correlation function that is expressed in comoving coordinates. It was later realized in~\cite{Senatore:2012nq,Gerstenlauer:2011ti,Giddings:2011zd,Giddings:2011ze} that they cancel completely out once the calculation is performed in local physical coordinates where the long-wavelength fluctuations are re-absorbed within the coordinates:  for example $V_{\rm physical}={\rm Exp}(3\zeta_L)a(t)^3 V_{\rm comoving}$. Since expressions (\ref{eq:long_short_average}) and (\ref{eq:final}) are in comoving coordinates, they are also affected by these IR divergencies, which are not harmful. More in detail, starting at 2-loops, the IR divergence (\ref{eq:expt2}) comes from the IR divergence of the short-mode correlation functions at horizon crossing. Notice that while for modes longer than the horizon the IR divergence is slow-roll suppressed, this is not the case for modes evaluated at horizon crossing. Indeed the three-point function in the squeezed limit is not slow-roll suppressed in this same kinematical regime~\cite{Pimentel:2012tw,Senatore:2012wy}. The second line of (\ref{eq:long_short_average}) is also IR divergent starting at 2-loops as it also involves short mode correlations in the background of long wavelength modes. These IR divergencies need to be present in correlation functions in comoving coordinates in order to ensure that once the correlation functions are expressed in terms of local physical coordinates, every IR dependence on modes much longer than the ones of interest is absent~\cite{Senatore:2012nq}.

\subsubsection*{Acknowledgments}

L.S.~is supported by by DOE Early Career Award DE-FG02-12ER41854 and the National Science Foundation under PHY-1068380.
M.Z. is supported by the National Science Foundation under PHY-0855425, AST-0907969 and
PHY-1213563 and by the David and Lucile Packard Foundation.

 \begingroup\raggedright\endgroup

%\bibliography{trapbib}
\end{document}